\documentclass{article}

\usepackage{color}
\usepackage{amsmath, amssymb}

\newcommand{\RM}{\mathbb{R}}
\newcommand{\ZM}{\mathbb{Z}}

\newcommand{\CM}{\mathbb{C}}
\newtheorem{theorem}{Theorem}
 
\newtheorem{prop}{Proposition} 
\newtheorem{cor}{Corollary}

\newcommand{\kvec}{\ensuremath{\boldsymbol{k}}}
\newcommand{\wvec}{\ensuremath{\boldsymbol{w}}}

\begin{document}

\title{{\bf CTM/Zeta Correspondence}
\vspace{15mm}}

\author{Takashi KOMATSU \\
Math. Research Institute Calc for Industry \\
Minami, Hiroshima, 732-0816, Japan \\ 
e-mail: ta.komatsu@sunmath-calc.co.jp 
\\ \\
Norio KONNO \\
Department of Applied Mathematics, Faculty of Engineering \\ 
Yokohama National University \\
Hodogaya, Yokohama, 240-8501, Japan \\
e-mail: konno-norio-bt@ynu.ac.jp \\
\\ 
Iwao SATO \\ 
Oyama National College of Technology \\
Oyama, Tochigi, 323-0806, Japan \\ 
e-mail: isato@oyama-ct.ac.jp 
}

\date{\empty }

\maketitle

\vspace{50mm}


\vspace{20mm}











\clearpage

\begin{abstract}
In our previous work, we investigated the relation between zeta functions and discrete-time models including random and quantum walks. In this paper, we introduce a zeta function for the continuous-time model (CTM) and consider CTMs including the corresponding random and quantum walks on the $d$-dimensional torus.
\end{abstract}

\vspace{10mm}

\begin{small}
\par\noindent
{\bf Keywords}: Zeta function, Continuous-time quantum walk, Continuous-time random walk, Torus
\end{small}

\vspace{10mm}

\section{Introduction \label{sec01}}
Our previous three papers \cite{KomatsuEtAl2021a, KomatsuEtAl2021b, KomatsuEtAl2021c} are devoted to the relation between zeta functions and {\it discrete-time} models (DTMs); Grover walk in \cite{KomatsuEtAl2021a}, random walk (RW) and quantum walk (QW) in \cite{KomatsuEtAl2021b}, interacting particle system (IPS) with probabilistic or quantum interactions in \cite{KomatsuEtAl2021c}. As for QW, see \cite{Konno2008, ManouchehriWang, Portugal, Venegas}, as for RW, see \cite{Konno2009, Norris, Spitzer}, and as for IPS, see \cite{Durett1988}, for examples. The corresponding works are called ``Grover/Zeta, Walk/Zeta, and IPS/Zeta Correspondence", respectively.

On the other hand, the present paper deals with the relation between zeta functions and {\em continuous-time} models (CTMs), in particular, continuous-time RW (CTRW) and continuous-time QW (CTQW). Concerning CTRW and CTQW, see \cite{Konno2005, Norris}. We call this relation ``CTM/Zeta Correspondence" here.

In fact, compared with the DTM case, the corresponding zeta function for the CTM was not known. Thus, the motivation of our study is to find out such a zeta function. Moreover, it is to clarify the relation between the zeta function for the DTM and that for the CTM.

The rest of this paper is organized as follows. Section \ref{sec02} is devoted to the definition of the CTM. In Section \ref{sec03}, we introduce a zeta function for the CTM and present our main result. Section \ref{sec04} deals with CTMs on the $d$-dimensional torus. In Section \ref{sec05}, we discuss the relation between our results and related topics. Finally, Section \ref{sec06} summarizes our results.

\section{CTM \label{sec02}}
First we introduce the following notation: $\mathbb{Z}$ is the set of integers, $\mathbb{Z}_{>} = \{1,2,3, \ldots \}$,  $\mathbb{R}$ is the set of real numbers, $\mathbb{R}_{\ge} = [0, \infty)$, and $\mathbb{C}$ is the set of complex numbers. 

Let $P^{(D,c)}$ denote an $n \times n$ transposed {\em stochastic matrix} (also called {\em transition matrix}) for a discrete-time Markov chain, where superscript $D$ stands for ``Discrete" and $c$ for ``classical". Then we define the evolution matrix at time $t$ for the corresponding continuous-time classical model by
\begin{align}
P^{(C,c)} _t = \exp \left\{ t \left( P^{(D,c)} - I_n \right) \right\} \quad (t \in \mathbb{R}_{\ge}),
\label{def01}
\end{align}
where superscript $C$ stands for ``Continuous" and $I_n$ is the $n \times n$ identity matrix. As a special case, this model contains CTRW. Moreover, if we consider the location of CTRW as a configuration, then we can treat IPS with a probabilistic interaction.

On the other hand, when $P^{(D,c)}$ is symmetric, the evolution matrix at time $t$ for the corresponding continuous-time quantum model is given by
\begin{align}
P^{(C,q)} _t = \exp \left\{ i t \left( P^{(D,c)} - I_n \right) \right\} \quad (t \in \mathbb{R}_{\ge}),
\label{def02}
\end{align}
where superscript $q$ stands for ``quantum" and $i=\sqrt{-1}$. Remark that $P^{(C,q)} _t$ is unitary for any $t \in \mathbb{R}_{\ge}$. This model contains CTQW as a special case. If we consider the location of CTQW as a configuration, then we can treat IPS with a quantum interaction.

The present paper introduces the following evolution matrix at time $t$ for the CTM (including the above mentioned two models for extreme cases) with a parameter $\xi$ as 
\begin{align}
P^{(C,\xi)} _t = \exp \left\{ e^{i \xi} \cdot t \left( P^{(D,c)} - I_n \right) \right\} \quad (\xi \in [0, \pi/2], \  t \in \mathbb{R}_{\ge}).
\label{def03}
\end{align}
In fact, if $\xi=0$, then Eq. \eqref{def03} becomes Eq. \eqref{def01} (classical case), if $\xi=\pi/2$, then Eq. \eqref{def03} becomes Eq. \eqref{def02} (quantum case).

\section{Zeta Function \label{sec03}}
Let $G_n$ denote a simple connected graph with $n$ vertices. Following our previous papers \cite{KomatsuEtAl2021a, KomatsuEtAl2021b, KomatsuEtAl2021c}, we introduce a {\em zeta function} for CTM on $G_n$ as  
\begin{align}
\overline{\zeta}^{(C,\xi)} _t \left(G_n, u \right) = \left\{ \det \Big( I_{n} - u P^{(C,\xi)} _t \Big) \right\}^{-1/n} \quad (\xi \in [0, \pi/2], \ t \in \mathbb{R}_{\ge}).
\label{satosan01}
\end{align}
Here, the variable $u$ is a sufficiently small complex number. Furthermore, we define $C^{(C,\xi)} _{r,n,t} (\in \CM)$ by
\begin{align}
\overline{\zeta}^{(C,\xi)} _t \left(G_n, u \right) 
= \exp \left( \sum_{r=1}^{\infty} \frac{C^{(C,\xi)} _{r,n,t}}{r} u^r \right).
\label{satosan03}
\end{align}
Let ${\rm Spec} ( A )$ denote the set of eigenvalues of a square matrix $A$. We put
\begin{align}
{\rm Spec} \left( P^{(D,c)} \right) = \left\{  \lambda_{j} \ | \ j = 1, 2, \ldots, n \right\}.
\label{koyuchiD} 
\end{align}

Then we have the following result.
\begin{theorem}
\begin{align*}
\overline{\zeta}^{(C,\xi)} _t \left(G_n, u \right) ^{-1}
&= \exp \left[ \frac{1}{n} \sum_{j=1}^{n} \log \Big( 1 - u \exp \left\{ e^{i \xi} \cdot t \left( \lambda_j - 1 \right) \right\} \Big) \right],
\\
C^{(C,\xi)} _{r,n,t}
&= \frac{1}{n} \sum_{j=1}^n \exp \left\{ e^{i \xi} \cdot r t \left( \lambda_j - 1 \right) \right\}.
\end{align*}
\label{thm001}
\end{theorem}
\par
\
\par\noindent
{\bf Proof.} Combining Eqs. \eqref{def03} and \eqref{satosan01} with Eq. \eqref{koyuchiD} implies 
\begin{align*}
\overline{\zeta}^{(C,\xi)} _t \left(G_n, u \right)^{-1}
&= \left\{ \det \Big( I_{n} - u P^{(C,\xi)} _t \Big) \right\}^{1/n} 
\\
&= \left\{ \det \Big( I_{n} - u \exp \left\{ e^{i \xi} \cdot t \left( P^{(D,c)} - I_n \right) \right\} \Big) \right\}^{1/n} 
\\
&= \left\{ \prod_{j=1}^n \Big( 1 - u \exp \left\{ e^{i \xi} \cdot t \left( \lambda_j - 1 \right) \right\} \Big) \right\}^{1/n} 
\\
&= \exp \left[ \frac{1}{n} \log \left\{ \prod_{j=1}^n \Big( 1 - u \exp \left\{ e^{i \xi} \cdot t \left( \lambda_j - 1 \right) \right\} \Big) \right\} \right]
\\
&= \exp \left[ \frac{1}{n} \sum_{j=1}^n \log \Big( 1 - u \exp \left\{ e^{i \xi} \cdot t \left( \lambda_j - 1 \right) \right\} \Big) \right].
\end{align*}
So we have the first part. As for the second part, from Eqs. \eqref{satosan01} and \eqref{satosan03}, we get 
\begin{align*}
\left\{ \det \Big( I_{n} - u P^{(C,\xi)} _t \Big) \right\}^{-1/n} = \exp \left( \sum_{r=1}^{\infty} \frac{C^{(C,\xi)} _{r,n,t}}{r} u^r \right).
\end{align*}
Thus 
\begin{align}
- \frac{1}{n} \log \left\{ \det \Big( I_{n} - u P^{(C,\xi)} _t \Big) \right\} = \sum_{r=1}^{\infty} \frac{C^{(C,\xi)} _{r,n,t}}{r} u^r.
\label{atami02D}
\end{align}
In a similar way, it follows from Eq. \eqref{koyuchiD} that the left-hand side of Eq. \eqref{atami02D} becomes
\begin{align*}
- \frac{1}{n} \log \left\{ \det \Big( I_{n} - u P^{(C,\xi)} _t \Big) \right\} 
&= - \frac{1}{n} \log \left\{ \prod_{j=1}^n \Big( 1 - u \exp \left\{ e^{i \xi} \cdot t \left( \lambda_j - 1 \right) \right\} \Big) \right\}
\\
&= - \frac{1}{n} \sum_{j=1}^n \log \Big( 1 - u \exp \left\{ e^{i \xi} \cdot t \left( \lambda_j - 1 \right) \right\} \Big)
\\
&= \frac{1}{n} \sum_{j=1}^n \sum_{r=1}^{\infty} \frac{\left( \exp \left\{ e^{i \xi} \cdot t \left( \lambda_j - 1 \right) \right\} \right)^r}{r} u^r. 
\end{align*}
By this and the right-hand side of Eq. \eqref{atami02D}, we have the desired conclusion:
\begin{align*}
C^{(C,\xi)} _{r,n,t}
= \frac{1}{n} \sum_{j=1}^n \exp \left\{ e^{i \xi} \cdot r t \left( \lambda_j - 1 \right) \right\}.
\end{align*}

\section{CTM on Torus \label{sec04}}
Let $T^d_N$ denote the {\em $d$-dimensional torus} where $d, N \in \mathbb{Z}_{>}$. Remark that $T^d_N$ is expressed as $(\mathbb{Z} \ \mbox{mod}\ N)^{d}$ and has $N^d$ vertices. 

Let 
\begin{align*}
e^{(n)} (\wvec) = e^{(n)} (w_1, w_2, \ldots, w_n) = w_1 + w_2 + \cdots + w_n
\end{align*}
for $\wvec = (w_1, w_2, \ldots, w_n) \in \CM^n$ and $n \in \ZM_{>}.$ Moreover we put $\mathbb{K}_N = \{ 0,1, \ldots, N-1 \}$ and $\widetilde{\mathbb{K}}_N = \{ 0 ,2 \pi/N, \ldots, 2 \pi (N-1)/N \}$. For $\kvec=(k_1,k_2,\ldots,k_{d}) \in \mathbb{K}_N^d$, we define 
\begin{align*}
\widetilde{k}_j = \frac{2 \pi k_j}{N} \in \widetilde{\mathbb{K}}_N, \quad \widetilde{\kvec}=(\widetilde{k}_1,\widetilde{k}_2,\ldots,\widetilde{k}_{d}) \in \widetilde{\mathbb{K}}_N^d.
\end{align*}
In this setting, we introduce
\begin{align*}
e^{(n, \cos)} (\widetilde{\kvec}) 
=  e^{(n)} (\cos \widetilde{k}_1, \cos \widetilde{k}_2, \ldots, \cos \widetilde{k}_n), 
\end{align*}
where $n \in \ZM_{>}$. Similarly, for $\Theta^{(n)} = (\theta_1, \theta_2, \ldots, \theta_n) (\in [0, 2 \pi)^n)$,  
\begin{align*}
e^{(n, \cos)} (\Theta^{(n)}) =  e^{(n)} (\cos \theta_1, \cos \theta_2, \ldots, \cos \theta_n),
\end{align*}
where $n \in \ZM_{>}$.

From now on, we consider CTMs on $T^d_N$ including CTRW and CTQW as special cases. From definition of the (simple symmetric) DTRW on $T^d_N$ (see \cite{Norris, Spitzer}), we easily see that 
\begin{align} 
{\rm Spec} \left( P^{(D,c)} \right) 
&= \left\{ \frac{1}{d} \sum^d_{j=1} \cos \left( \frac{2 \pi k_j }{N} \right) \bigg| \ k_1 , \ldots , k_d \in \mathbb{K}_N \right\} 
\nonumber
\\
&= \left\{ \frac{1}{d} e^{(d, \cos)} (\widetilde{\kvec}) \bigg| \ k_1 , \ldots , k_d \in \mathbb{K}_N \right\}.
\label{specP}
\end{align}
Here the DTRW on $T^d_N$ jumps to each of its nearest neighbours with equal probability $1/2d$.

Let $d \Theta^{(d)}_{unif}$ denote the uniform measure on $[0, 2 \pi)^d$, that is,
\begin{align*}
d \Theta^{(d)}_{unif} = \frac{d \theta_1}{2 \pi } \cdots \frac{d \theta_d}{2 \pi }.
\end{align*}

Applying Eq. \eqref{specP} to Theorem \ref{thm001} gives
\begin{cor}
\begin{align*}
\overline{\zeta}^{(C,\xi)} _t \left(T^d_N, u \right) ^{-1}
&= \exp \Bigg[ \frac{1}{N^d} \sum^{N-1}_{ k_1 =0} \cdots \sum^{N-1}_{ k_d =0} \log \Bigg\{ F^{(C,\xi)} \left( \widetilde{\kvec}, u \right) \Bigg\} \Bigg],
\\
\lim_{N \to \infty} \overline{\zeta}^{(C,\xi)} _t \left(T^d_N, u \right) ^{-1}
&= \exp \Bigg[ \int_{[0,2 \pi)^d} \log \Bigg\{ F^{(C,\xi)} \left( \Theta^{(d)}, u \right) \Bigg\} d \Theta^{(d)}_{unif} \Bigg],
\end{align*}
where 
\begin{align*}
F^{(C,\xi)} \left( \wvec, u \right) = 1 - u \times \exp \left\{ e^{i \xi} \cdot t \left( \frac{1}{d} e^{(d, \cos)} (\wvec) - 1 \right) \right\}.
\end{align*}
\label{cor001}
\end{cor}
 We should note that when we take a limit as $N \to \infty$, we assume that the limit exists throughout this paper.

Let $I_{\alpha} (x)$ be the modified Bessel function with the first kind of order $\alpha$ as
\begin{align*}
I_{\alpha} (x) = \sum_{k=0}^{\infty} \frac{\left(x/2 \right)^{2k+\alpha}}{k! \Gamma (\alpha + k +1)},
\end{align*}
where $\Gamma (x)$ is the gamma function (see Andrews et al. \cite{Andrews1999}, Watson \cite{Watson1944}). In a similar fashion, applying Eq. \eqref{specP} to Theorem \ref{thm001} yields 
\begin{cor}
\begin{align}
C^{(C,\xi)} _{r,N^d,t}
&= \frac{1}{N^d} \sum^{N-1}_{ k_1 =0} \cdots \sum^{N-1}_{ k_d =0} G^{(C,\xi)} \left( \widetilde{\kvec} \right),
\nonumber
\\
\lim_{N \to \infty} C^{(C,\xi)} _{r,N^d,t}
&= \int_{[0,2 \pi)^d} G^{(C,\xi)} \left( \Theta^{(d)} \right) d \Theta^{(d)}_{unif} 
\nonumber
\\
&
= \exp \left( - e^{i \xi} \cdot r t \right) \cdot \left\{ I_0 \left( \frac{e^{i \xi} \cdot r t}{d} \right) \right\}^d,
\label{doseki}
\end{align}
where 
\begin{align*}
G^{(C,\xi)} \left( \wvec \right) = \exp \left\{ e^{i \xi} \cdot r t \left( \frac{1}{d} e^{(d, \cos)} (\wvec) - 1 \right) \right\}.
\end{align*} 
\label{cor002}
\end{cor}

In the final part of this section, following the discrete-time walk case in our previous paper \cite{KomatsuEtAl2021b}, we consider DTM on a graph $G_n$ with $n$ vertices. Then the zeta function for DTM is given by
\begin{align}
\overline{\zeta}^{(D,c)} \left(G_n, u \right) = \left\{ \det \Big( I_{n} - u P^{(D,c)} \Big) \right\}^{-1/n}.
\label{satosan01D}
\end{align}
Recall that the zeta function for CTM is 
\begin{align*}
\overline{\zeta}^{(C,\xi)} _t \left(G_n, u \right) = \left\{ \det \Big( I_{n} - u P^{(C,\xi)} _t \Big) \right\}^{-1/n} \quad (\xi \in [0, \pi/2], \ t \in \mathbb{R}_{\ge}),
\end{align*}
where 
\begin{align*}
P^{(C,\xi)} _t = \exp \left\{ e^{i \xi} \cdot t \left( P^{(D,c)} - I_n \right) \right\}.
\end{align*}

Moreover, we define $C^{(D,c)} _{r,n} (\in \CM)$ by
\begin{align}
\overline{\zeta}^{(D,c)} \left(G_n, u \right) 
= \exp \left( \sum_{r=1}^{\infty} \frac{C^{(D,c)} _{r,n}}{r} u^r \right).
\label{satosan03D}
\end{align}
As in the proof of Theorem \ref{thm001}, from Eqs. \eqref{koyuchiD}, \eqref{satosan01D}, and \eqref{satosan03D}, we have the following result.
\begin{prop}
\begin{align*}
\overline{\zeta}^{(D,c)} \left(G_n, u \right) ^{-1}
&= \exp \left[ \frac{1}{n} \sum_{j=1}^{n} \log \Big( 1 - u \lambda_j \Big) \right],
\\
C^{(D,C)} _{r,n}
&= \frac{1}{n} \sum_{j=1}^n (\lambda_j)^r.
\end{align*}
\label{atami03D}
\end{prop}
Furthermore, we consider DTRW on $d$-dimensional torus $T^d_N$. In \cite{KomatsuEtAl2021b}, we treated DTRW only on one-dimensional torus $T^1_N$. So in order to clarify the difference between CTRW and DTRW on $T^d_N$, we will deal with general $d$-dimensional case. Applying Eq. \eqref{specP} to Proposition \ref{atami03D} implies
\begin{cor}
\begin{align*}
\overline{\zeta}^{(D,c)} \left(T^d_N, u \right) ^{-1}
&= \exp \Bigg[ \frac{1}{N^d} \sum^{N-1}_{ k_1 =0} \cdots \sum^{N-1}_{ k_d =0} \log \Bigg\{ F^{(D,c)} \left( \widetilde{\kvec}, u \right) \Bigg\} \Bigg],
\\
\lim_{N \to \infty} \overline{\zeta}^{(D,c)} \left(T^d_N, u \right) ^{-1}
&= \exp \Bigg[ \int_{[0,2 \pi)^d} \log \Bigg\{ F^{(D,c)} \left( \Theta^{(d)}, u \right) \Bigg\} d \Theta^{(d)}_{unif} \Bigg],
\end{align*}
where 
\begin{align*}
F^{(D,c)} \left( \wvec, u \right) = 1 - u \times \frac{1}{d} e^{(d, \cos)} (\wvec).
\end{align*}
Moreover, we have
\begin{align*}
C^{(D,c)} _{r,N^d}
&= \frac{1}{N^d} \sum^{N-1}_{ k_1 =0} \cdots \sum^{N-1}_{ k_d =0} G^{(D,c)} \left( \widetilde{\kvec} \right),
\\
\lim_{N \to \infty} C^{(D,c)} _{r,N^d}
&= \int_{[0,2 \pi)^d} G^{(D,c)} \left( \Theta^{(d)} \right) d \Theta^{(d)}_{unif} ,
\end{align*}
where 
\begin{align*}
G^{(D,c)} \left( \wvec \right) = \left( \frac{1}{d} e^{(d, \cos)} (\wvec) \right)^r.
\end{align*} 
\label{cor002D}
\end{cor}
Remark that $\lim_{N \to \infty} C^{(D,c)} _{r,N^d}$ is nothing but the return probability for DTRW at time $r$. In fact, if $d=1$ and $d=2$, then
\begin{align*}
\lim_{N \to \infty} C^{(D,c)} _{r,N}
&= \int_{0}^{2 \pi} \left( \cos \theta \right)^r \frac{d \theta}{2 \pi} 
= \left\{ 
\begin{array}{ll}
{\displaystyle {r \choose r/2} \left( \frac{1}{2} \right)^r } & \mbox{if $r$ is even, } 
\\
\\
0 & \mbox{if $r$ is odd},
\end{array}
\right.
\\
\lim_{N \to \infty} C^{(D,c)} _{r,N^2} 
&= \left( \lim_{N \to \infty} C^{(D,c)} _{r,N} \right)^2.
\end{align*}
However, if $d = 3, 4, \ldots$, then such a simple form is not known (see \cite{Norris, Spitzer}).

\section{Discussion \label{sec05}}
In this section, we discuss the relation between our results and related topics. For the sake of simplicity, we focus on the one-dimensional case $\ZM$. First we define the discrete {\em Laplacian} $\Delta$ by 
\begin{align*}
\Delta F (x) = F (x-1) + F (x+1) - 2 F (x) \qquad (x \in \ZM).
\end{align*}
By using this, we consider the following {\em diffusion equation}:
\begin{align}
\frac{\partial \phi (t,x)}{\partial t} = \frac{1}{2} \Delta \phi (t,x) \qquad \left( t \in \RM_{\ge}, \ x \in \ZM \right).
\label{dCTM}
\end{align}
Moreover we introduce the fundamental solution:
\begin{align*}
g_t (x) = e^{-t} \ I_{x} (t) \qquad \left( t \in \RM_{\ge}, \ x \in \ZM \right),\end{align*}
where $I_{\alpha} (x)$ is the modified Bessel function with the first kind of order $\alpha$.  
For an initial condition $\phi (0,x) = f(x)$, where $f: \ZM \to \RM$ is a suitable function, a solution of the diffusion equation \eqref{dCTM} is given by  
\begin{align*}
\phi (t,x) 
&= E \left[ f \left( x + S_t \right) \right] 
= \sum_{y \in \ZM} f (x+y) \ P \left( S_t = y \right)
\\
&= \sum_{y \in \ZM} f (x+y) \ e^{-t} \ I_{y} (t)
= (g_t \ast f) (x),
\end{align*}
where $S_t$ is CTRW starting from the origin on $\ZM$ and $\ast$ means the convolution. We should note that
\begin{align*}
P (S_t = x) = g _{t} (x) = e^{-t} \ I_{x} (t) \qquad \left( t \in \RM_{\ge}, \ x \in \ZM \right).
\end{align*}
Next we consider the following {\em Schr\"odinger equation}: 
\begin{align*}
i \ \frac{\partial \Psi (t,x)}{\partial t} = - \frac{1}{2} \Delta \Psi (t,x) \qquad \left( t \in \RM_{\ge}, \ x \in \ZM \right).
\end{align*}
This is rewritten as  
\begin{align}
\frac{\partial \Psi (t,x)}{\partial t} = i \times \frac{1}{2} \Delta \Psi (t,x) \qquad \left( t \in \RM_{\ge}, \ x \in \ZM \right).
\label{sCTM}
\end{align}
As in the case of the above mentioned diffusion equation \eqref{dCTM}, for an initial condition $\phi (0,x) = f(x)$, where $f: \ZM \to \CM$ is a suitable function, a solution of the Schr\"odinger equation \eqref{sCTM} is given by  
\begin{align*}
\Psi (t,x) = (g_{it} \ast f) (x).
\end{align*}
Remark that
\begin{align*}
g _{it} (x) = e^{-it} \ I_{x} (it) = e^{-it} \ i^{x} \ J_{x} (t) \qquad \left( t \in \RM_{\ge}, \ x \in \ZM \right),
\end{align*}
where $J_{\alpha} (x)$ denotes the Bessel function with the first kind of order $\alpha$ as 
\begin{align*}
J_{\alpha} (x) = \sum_{k=0}^{\infty} \frac{(-1)^k\left(x/2 \right)^{2k+\alpha}}{k! \Gamma (\alpha + k +1)}.
\end{align*}

Let $X_t$ be CTQW starting from the origin on $\ZM$. Like CTRW on $\ZM$, we see that
\begin{align*}
P (X_t = x) = |g _{it} (x)|^2 = |I_{x} (it)|^2 = J_{x} (t)^2 \qquad \left( t \in \RM_{\ge}, \ x \in \ZM \right).
\end{align*}
As for CTRW and CTQW, see \cite{Konno2005, Norris}, for example. In this setting, our CTM model is related to the following partial differential equation:
\begin{align}
\frac{\partial \psi^{(\xi)} (t,x)}{\partial t} = e^{i \xi} \times \frac{1}{2} \Delta \psi^{(\xi)} (t,x) \qquad \left( \xi \in [0, \pi/2], \ t \in \RM_{\ge}, \ x \in \ZM \right).
\label{eCTM}
\end{align}
If $\xi =0$, then Eq. \eqref{eCTM} becomes Eq. \eqref{dCTM} (diffusion equation) related to CTRW on $\ZM$, if $\xi=\pi/2$, then Eq. \eqref{eCTM} becomes Eq. \eqref{sCTM} (Schr\"odinger equation) related to CTQW on $\ZM$. As in the cases of the above mentioned diffusion equation \eqref{dCTM} and Schr\"odinger equation \eqref{sCTM}, for an initial condition $\psi^{(\xi)} (0,x) = f(x)$, where $f: \ZM \to \CM$ is a suitable function, a solution of Eq. \eqref{eCTM} is given by  
\begin{align*}
\psi^{(\xi)} (t,x) = (g_{e^{i \xi} \cdot t} \ast f) (x).
\end{align*}
Here
\begin{align}
g_{e^{i \xi} \cdot t} (x) = e^{- e^{i \xi} \cdot t} \ I_{x} (e^{i \xi} \cdot t) \qquad \left( \xi \in [0, \pi/2], \ t \in \RM_{\ge}, \ x \in \ZM \right).
\label{dosekikowai}
\end{align}
We should note that Eq. \eqref{dosekikowai} with $x=0$ (the origin) is equal to $\lim_{N \to \infty} C^{(C,\xi)} _{r,N^d,t}$ in Eq. \eqref{doseki} with $r=1$ and $d=1$.

\section{Summary \label{sec06}}
In our previous works \cite{KomatsuEtAl2021a, KomatsuEtAl2021b, KomatsuEtAl2021c}, we studied the relation between zeta functions and DTMs including Grover walk, DTRW, DTQW, and IPS, as special cases. The corresponding works are called ``Grover/Zeta, Walk/Zeta, and IPS/Zeta Correspondence", respectively. On the other hand, this paper introduced a zeta function for CTM and investigated the relation between the zeta function and CTMs including CTRW and CTQW as extreme cases. We called this relation ``CTM/Zeta Correspondence". In particular, we treated CTMs on the $d$-dimensional torus, so one of the interesting future problems might be to extend the torus to a suitable class of graphs.


\end{document}